\documentstyle[prd,preprint,aps]{revtex}

\newcommand{\be}{\begin{equation}}
\newcommand{\ee}{\end{equation}}
\newcommand{\bea}{\begin{eqnarray}}
\newcommand{\eea}{\end{eqnarray}}

\begin{document}

\reversemarginpar
\tighten

\title{A comment on  black hole entropy \\ or \\  does 
Nature abhor a logarithm?} 

\author {A.J.M. Medved}

\address{
School of Mathematical and Computing Sciences\\
Victoria University of Wellington\\
PO Box 600, Wellington, New Zealand \\
E-Mail: joey.medved@mcs.vuw.ac.nz}

\maketitle

\begin{abstract}

There has been substantial interest, as of late, in the 
 quantum-corrected form of  the Bekenstein--Hawking black hole entropy. 
The consensus viewpoint is that the leading-order correction
should be a logarithm of the horizon area; however, the value
of the logarithmic prefactor remains a point of notable controversy.
Very recently, Hod has employed statistical arguments that
constrain this prefactor to be a non-negative integer.
In the current paper, we invoke some  independent considerations
to argue that the ``best guess'' for the prefactor might simply be
zero.  Significantly, this value complies with the prior prediction and,
moreover,  seems  suggestive of some  fundamental symmetry.

\end{abstract}

\section{Motivation}

It has long been accepted that black holes possess
 an intrinsic  entropy which (for at least a wide class
of models) can be determined by way
of the famous Bekenstein--Hawking area law \cite{bek1,hawk}; that is,
\be
S_{BH}={A\over 4}\;,
\label{1}
\ee
where $S_{BH}$ is the entropy in question and $A$
is the cross-sectional area (in Planck units) of the black hole
horizon.
Here and throughout,  all  fundamental
constants are set equal to unity. 
Moreover, we will  focus on  the physically realistic  case of
only four uncompactified spacetime dimensions, as well as
the case of a black hole that is neutral and static {\it modulo} quantum
fluctuations
(although many
of our statements  have more general applicability).
Also,
we will always assume the semi-classical regime of a macroscopically large
 black hole or  $A>>1$. Further note that the above relation is,
in spite of the implied presence of $\hbar$ (in the denominator),  
strictly a classical one.

Although
the black hole  area law  
initially followed from
 thermodynamic considerations ({\it e.g.}, protecting
the second law of thermodynamics in the presence of a black
hole \cite{bek1}), it is often presumed to
have a statistical meaning as well \cite{carlip}.
Which is to say, one would naturally expect there to be a (yet-to-be-clarified)
set of microscopic states that can account for this entropy
by way of state-counting procedures. At this point, it is reasonable
to suggest that such a statistical framework will ultimately be traced
to some fundamental theory that is able to
unify gravitational physics with  quantum mechanics. 

With cognizance of the above ideas, it becomes apparent that
the Bekenstein--Hawking area law will provide an
important viability test for any prospective theory of
quantum gravity \cite{smo}.  In fact, there has already been remarkable
success in this direction:  statistical
calculations of the black hole entropy in the context of (for instance) loop
quantum gravity \cite{asht}, string theory \cite{strv}, 
``Sakharov-style'' induced  
gravity \cite{fro},  near-horizon symmetries \cite{solod}, 
{\it etcetera}.  Perhaps, even too much success
because, clearly, some  further discrimination is still required.
This realization has helped spark some recent interest
 into calculating  the leading-order quantum correction to 
the classical quantity, $S_{BH}$ \cite{kamaj}. 

In the remainder, we will elaborate on the nature of this correction
and --- with input from a variety of sources --- make a
novel deduction about what could be its precise value.  We will end
with a discussion on the possible implications (followed by
an addendum).

\section{Quantum-Corrected Entropy}

Let us begin here by taking note of the typical outcome for the
quantum-corrected entropy, $S_{bh}$,
as obtained
by a variety of methodologies:~\footnote{For a
thorough list of references, see  \cite{setare}.} 
\be 
S_{bh}=S_{BH}+b\ln[S_{BH}] +\; \rm{constant}\;+{\cal O}[A^{-1}]\; .
\label{2}
\ee
That is, the leading-order correction is, almost universally,
found to be a logarithm of  the horizon area. Unfortunately,
there has been no such consensus with regard to the 
value of the logarithmic prefactor --- denoted here by $b$.
In fact,  from a survey of the literature, the prefactor appears 
to be  a highly  model and method dependent 
parameter.  On the other hand, as will be  detailed later on,
much of this ambiguity in  $b$ can be attributed
to the following (often overlooked) point of relevance:
There happens to be two distinct and separable sources 
to this logarithmic correction \cite{gg2,chatt}. Nevertheless, even
equipped  with this knowledge, it has   remained  unclear as to how
one might unambiguously fix this parameter. (A partially conjectural
argument  will, however, be presented in due course.)

Supposing that the value of $b$ cannot be fixed, one might
ask if there is any means for at least constraining  this prefactor.  
As it so happens, this very question
was recently addressed in a paper by Hod \cite{hod}.~\footnote{The
idea of constraining this parameter, 
but in the context of loop quantum gravity,
has also been  considered in  \cite{mitra}.} 
The premise of this work was to take seriously the
statistical interpretation of the entropy in the context
of Bekenstein's notion of a  quantum black hole 
area spectrum \cite{bek2}. [The latter notion being
that the horizon area, $A$, should be  quantized according to 
\be
A_n=\epsilon  n \;,
\label{3}
\ee
where $n=0,1,2,...$ and $\epsilon$ is a numerical
factor of the order unity. ($A_n$ might also
include a constant term or zero-point area.) 
There has been substantial work in support of
this form of spectrum \cite{bekg}; most simply, it follows from
the horizon area being an adiabatic invariant \cite{bekx}  along with 
Ehrenfest's principle.] 

More precisely, ``in the spirit of the Boltzmann--Einstein formula
in statistical physics'' \cite{hod}, Hod suggested that the number
of horizon 
microstates  which correspond to a particular macrostate
should be fixed equal to a positive integer.~\footnote{A similar
point of view was used in Hod's earlier conjecture; which essentially
identified  the spacing  of Bekenstein's area spectrum with 
the real part of the quasinormal mode frequency of 
a highly damped black hole \cite{hod2}.} Given this line of reasoning,
the degeneracy of the $n$th area eigenvalue (which translates into
 the exponential of the entropy)
should be  constrained as follows \cite{hod}:
\be
g_n \equiv \exp[S_{bh}(n)]= \;{\rm positive\; integer}\;,
\label{4}
\ee
for {\it any} allowed value of $n$.
It should be noted that this constraining relation does not depend
on what choice is made for the area spectrum [{\it i.e.}, at
this point, $A_n$ may or may not comply with equation (\ref{3}).]

Hod then proceeded by writing  the quantum-corrected 
entropy and area spectra in the most general
forms that are consistent with the classical area law.
More explicitly, $A_n$ was written as an expansion in terms of $n$,
and  $S_{bh}$ was expanded  in terms 
of $A$;  where both expansions include a linear term,
logarithmic term, a constant, as well as  any number of  sub-leading 
power-law terms 
(with the latter inclusion turning out to be irrelevant).
To comply with the constraint
equation (\ref{4}), it was then shown that the 
entropy expansion must necessarily reduce to the form \cite{hod}
\be 
S_{bh}=S_{BH}+l\ln[S_{BH}] + \ln\left[{(m+1)4^l\over\epsilon^{l}}\right]\;, 
\label{5}
\ee 
where $\epsilon$ is the coefficient of the linear term in the area
expansion, and both $l$ and $m$ are non-negative integers.~\footnote{Here,
we have slightly modified the notation of \cite{hod}.}
As a further point of interest, the area spectrum must
reduce precisely to the form   of equation (\ref{3}) [except
for the viable possibility of a $\ln(n)$ term, but  only
if   $l=0$ in equation (\ref{5})].

Comparing the above  outcome   with the
quantum-corrected  entropy of 
equation (\ref{2}),  we see that the logarithmic prefactor,
or $b$, has now been  constrained to be a non-negative integer.
(Furthermore, the constant  term has been somewhat constrained
and the lower-order corrections can now be safely  dropped.)
The question we are now inclined to ask is how does
such a condition on $b$ comply with what is known (or can be deduced) from  
previous studies?

Before directly addressing this last query, let us first consider
the possible sources for this logarithmic correction
to the classical entropy. As alluded to above, there are
two distinct contributing 
factors that need to be accounted for \cite{gg2,chatt}. Firstly, there may 
be a quantum correction to the {\it microcanonical} entropy;
that is, a fundamental correction to the number of microstates
that are required to describe a black hole of fixed
horizon area (and, consequently, fixed energy.) On intuitive
grounds,  a process of quantization should tend to  remove uncertainty about
the system, which  would then  typically   translate into a
decrease in the entropy.
On the other hand, without a full understanding of the
microcanonical degrees of freedom, it is impossible
to say definitively that this would be the case. Nevertheless,
if we assume that quantizing the black hole does 
reduce its entropy (an assumption that is supported by various
calculations in the literature --- see below), 
the  microcanonical contribution to
$b$ will then be negative. 
(Recall that $S_{BH}$ is always considered to be a very
large number.)   
 
Secondly, there may  be quantum  corrections
that arise due to the black hole exchanging heat with its surroundings.
(It is implicit that the black hole has been emersed in a thermal bath
of radiation; precisely  at the Hawking temperature. This is necessary 
for a state of thermal equilibrium to be reached; thereby fixing
the {\it mean} or classical  values of the energy and 
area.~\footnote{More accurately,
this state of affairs will fix the quantum-expectation values of
these quantities, while  still permitting
quantum fluctuations.})
These thermal fluctuations in the horizon area are best interpreted
as a {\it canonical} correction to the classical entropy.
Inasmuch as thermal fluctuations  increase the uncertainty of the system,
any  canonical contribution to $b$ must certainly be positive.
Hence, we have two conflicting contributions to the overall sign of
$b$, and it is not {\it a priori} clear which will win out.
Nonetheless, we will now proceed to argue (independently of
the preceding analysis) that the resultant
sum will indeed be a non-negative integer; in fact, it
turns out to be zero.

Before proceeding on with the main discussion, let us point out that Hod's
paper \cite{hod} considers the quantum corrections 
at the most fundamental level; thus implying that  the predictions therein 
refer
strictly to the microcanonical corrections. So, under the premise
of a comparison with \cite{hod}, one might ask
if it is fair to include canonical corrections in our calculation.
We will argue here that the canonical corrections should indeed
be included on the following basis:  The entropy of a physical system
is typically an observer-dependent concept, and this certainly applies
to a black hole, inasmuch as a free-falling observer
would detect no horizon (and, consequently, no associated entropy) whatsoever.
This means that, if we are to assign an operational meaning to the entropy,
a measurement must be performed or, equivalently, some sort of
interaction with the black hole is required. From this point of view
({\it i.e.}, the black hole interacting with its surroundings),
a canonical treatment would seem to be the appropriate one.

First of all, let us focus on  the microcanonical correction.
In principle, such a correction can only be properly quantified
at the  level of the (yet unknown) fundamental  gravity theory. Nevertheless,
certain calculations, especially in the conceptual framework
of {\it loop quantum gravity} \cite{kamaj}, have found a 
logarithmic correction
of $-{3\over 2}\ln S_{BH}$; that is, $b_{mc}=-3/2$ (in hopefully
obvious notation).~\footnote{It is worth pointing out
that this (loop quantum gravity)  calculation
does not seem to be tied to any particular gauge group \cite{kaul},
which is a good thing in view of some  recent controversies \cite{dreyer}.}
Actually, this finding is not really
as restrictive (in the sense of being tied to a specific
model of quantum gravity) 
as it might appear to be.   If the
quantization of the horizon  can be effectively described
in terms of an ``it from bit'' ({\it ala} Wheeler \cite{wheel}) model,
then one will be able to reproduce the same result.  To clarify somewhat,
the horizon surface is  segregated into patches
of Planckian size, with each  such patch represented
by the simplest possible operator; namely, 
 a single spin variable with spin equal to one half.
Hence, any of the patches is to be labeled with either ``spin up'' or  
``spin down''.~\footnote{The philosophical sector
of the audience may wish to replace the spin variable
with a {\it proposition} and  replace  up (down) 
with {\it true} ({\it false}) \cite{ish}.} 
Given this picture, a series of straightforward calculations reveals
that the entropy (defined as the logarithm  of the number of states
when the {\it sum} of the spins is zero)
reproduces both the area law and the advertised logarithmic correction
({\it i.e.}, $b_{mc}=-3/2$) \cite{khrip}. Even if this all feels
 rather {\it ad hoc},
the same basic calculation emerges quite
 naturally \cite{gg2} out of the algebraic approach to
 black hole quantization \cite{bekx}.  

When all is said and done, the most compelling argument for 
$b_{mc}=-3/2$ might be from a paper by Carlip \cite{carlip2}.
An essential ingredient in this calculation was 
 the leading-order correction to   Cardy's  entropic formula  
\cite{cardy}; notably,
 a formula that is applicable
to  a two-dimensional conformal
field theory.  Significantly, there is strong evidence 
that this type of field theory
can   effectively describe 
the horizon microstates of almost any stationary black hole 
\cite{carlip,solod}. 

We are now left with the daunting task of estimating the canonical
contribution to the correction. In this regard, let us first turn to
some  relevant papers by  Gour and the current author \cite{NEW},
and Chatterjee and Majumdar \cite{cmaju}.~\footnote{For
an earlier and a later example of related work, see \cite{kastrup} and
\cite{mup}.}  
In these treatments, the black hole is assumed (as noted above)  to be
in a state of thermal equilibrium with its surroundings;
for descriptive purposes, one might envision a black
hole enclosed in a ``reflective box''. [Pragmatically speaking,
one is able to mimic the scenario of  a Schwarzschild black hole in a box
by adopting the model of an anti-de Sitter (AdS) Schwarzschild black hole. 
This follows
from the AdS spacetime having a confining potential that acts
like a box of size $L$ \cite{hub}, 
where $L$ is the AdS radius of curvature.]
By virtue of this framework, the black hole system can now 
be effectively  modeled as a canonical ensemble of particles and fields.
The formal analysis \cite{NEW,cmaju} 
appropriately begins with a  partition function of the 
following form: 
\be
{\cal Z}_C(\beta,L)=\sum_{n}g_n \exp[\beta E(n;L)]\;,
\label{6}
\ee
where $\beta$ is the  temperature of the heat bath,
$n$ and $g_n$ are (as before) the quantum number and degeneracy
 for the area spectrum (with the latter, as usual, being identified with
the exponent of the microcanonical entropy), 
$E$ is the black hole energy or mass, and $L$ is the box size
or AdS radius when appropriate.

The calculation of the quantum-corrected entropy ({\it sans} microcanonical
corrections) can then be accomplished
in a number of steps. To briefly summarize:  \\
({\it i})   
the sum in equation (\ref{6}) 
is re-expressed  as an integral via the Poisson resummation formula,
\\ 
({\it ii}) the integration variable is changed to $E$
(taking into consideration a critical ``Jacobian'' factor), \\
({\it iii}) a saddle-point approximation is used to obtain
a Gaussian integral that can  readily be  evaluated, \\
({\it iv})
standard thermodynamic relations are then  utilized to
extract the canonical entropy from the partition function. 

Before proceeding any further, let us point out an
important caveat: One finds that,  in the next-to-last  step, 
the formalism dramatically  breaks down for any black
hole with a negative specific heat. Perhaps disturbingly, this would be 
the case
for even an ``ordinary'' Schwarzschild black hole. But, actually,
 this is not much of 
a surprise, given that a negative specific heat is
a sure  sign of thermodynamic instability.  Moreover, it is not 
 particularly worrisome, insofar as the specific heat
will remain positive for an AdS--Schwarzschild black hole
whenever $L$ is (roughly speaking) no larger than the
horizon size \cite{haw4}. That is,
the  ``size of the box'' must be sufficiently limited
 for the notion of a thermally equilibrated  black hole  to make 
any sort of sense. 

To finish off the outlined  calculation, the authors of \cite{cmaju}
assumed (along with the tree-level validity of the Bekenstein--Hawking 
entropy)  that the energy
has a simple power-law relation  (of arbitrary power) with
the area. [To be precise, this
assumption translates into the requirement  that
 $E\sim A^{\eta}$  is a valid leading-order expression
 for some choice of $\eta$.
Certainly, this  is  not particularly restrictive, although there
is a caveat which will be dealt with below.]
Ultimately, they  
found  a ``universal'' \cite{cmaju} 
logarithmic
correction (due to canonical effects) of the form 
 $+{1\over 2}\ln S_{BH}\;$; 
that is, $b_{C}=+1/2\;$.~\footnote{Note that the exact same result
was obtained in \cite{NEW}, although the model-specific focus of
that paper  shrouded the  universality of the calculation.}

Although we have reiterated the claim of ``universality'',
 this calculation is not
quite as general as it may  appear to be.
Besides maintaining a positive specific heat, the black hole system
is required  (given the condition of
 thermal stability) to  
 remain sufficiently far away
from any phase-transition points \cite{NEW,next}.  Furthermore,
it is in this very ``grey area'' that the assumed energy-area relation
will become invalid (due to two or more evenly competing terms when $E$ is 
expressed as a function of $A$). 
In Schwarzschild--AdS language, this
translates into a lower  bound on  the horizon radius (or, alternatively,
an upper limit on $L$) 
  that is somewhat larger 
than the bound set by the point of the Hawking--Page phase transition
or $A^{1/2}\sim L$ \cite{haw4}. [Actually,  the deviation between the
two bounds goes as  $L^{1/3}$ \cite{next}, which can be quite large
in Planck units, but is rather  small in relation to the size of a thermally
stable black hole.]
However, if
such thermally unstable regions are indeed avoided, then
any subleading terms in 
the  energy-area relation must necessarily show
up at lower than logarithmic order in the entropy. 
That is to say, given the validity of the canonical analysis,
this outcome is exact and universal within the regime of interest.

As it now stands, the canonical correction  would appear to
be inadequate to (at the very least)    cancel off 
the microcanonical correction; this being a minimal requirement
for compatibility with the Hod prediction.
However, the above cannot  be the whole story,
as we have not yet accounted for fluctuations
in the black hole  electrostatic charge and angular momentum.
Which is to say, even if the black hole is, itself,
neutral and static (by which we mean the equilibrium or classical values
of the charge and spin  are vanishing), the horizon area can
still  experience fluctuations due to a fluctuating  charge or 
spin.  It should be clear that  such fluctuations cannot be suppressed for any 
realistic form of  heat bath.

The pertinent question  now becomes  how might we quantitatively incorporate
the charge and spin fluctuations
into the canonical  correction?  Fortunately, for the case of a dynamical
 charge (we will momentarily neglect the issue of spin),
this calculation  has already been accomplished in \cite{NEW}. The 
key point is that
the formalism should now be  appropriately
generalized to that of a {\it grand-canonical}
ensemble. More specifically, a suitably revised form for
 the partition function is  expressible as follows \cite{NEW}:
\be
{\cal Z}_G(\beta,L)=\sum_{m}\sum_{n}g_{n,m} \exp[\beta E(n,m;L)-\mu Q(m)]\;,
\label{7}
\ee
where $m$ is a ``new'' quantum number that quantifies the black hole
charge ($Q=me\;$, where $e$ is a fundamental charge unit) 
and $\mu$ is an electric (or, more generally, chemical) 
potential. 
After some well-motivated choices for the level degeneracy
($g_{n,m}$) 
 and some lengthy calculations, it was shown \cite{NEW} that this  
remodeled 
framework ``typically'' yields  a logarithmic correction of 
$+1\ln S_{BH}\;$; meaning that, once the charge fluctuations
have been ``turned on'',   $b_C$ increases from $+1/2$
to $+1\;$.
Let us now clarify the range of validity or what is meant
by the qualifier ``typically'': 
This outcome of $b_C=+1$ will persist  as long as 
we (once again) assume a positive specific heat and
a sufficient separation from any phase-transition point,~\footnote{For
the case of an AdS--Reissner--Nordstrom black hole ---  a  natural
generalization of the AdS--Schwarzschild model --- 
there exists two such phase-transition points. One is the  Hawking--Page
 analogue \cite{haw4} and the other occurs  when both the charge and the box
are relatively large \cite{davies,next}.}
 {\it and}
(additionally) assume
a small enough charge.  In particular, the black hole cannot be
too close to extremality, where {\it all} the fluctuations would be 
suppressed \cite{NEW}.
But note that, for the physically realistic case of a black hole that
is at (or  close to)  a state of classical neutrality --- as has assumed 
to be the case here ---  these are precisely the same
conditions  as discussed earlier.

Let us re-emphasize that, assuming  the regime of thermal stability, 
 the current analysis is independent
of whether or not the black hole is classically neutral.
For either eventuality, one can still anticipate large fluctuations
in the charge  on account of 
the continual ebb and flow
of charged particles between  the black hole and  thermal
bath. 
In fact,
it has been shown that, even for a classically neutral black hole,
 the charge fluctuations go as $\Delta Q\sim \Delta A\sim S_{BH}^{1/2}$
\cite{NEW}.~\footnote{Given classical neutrality,
the {\it relative} size of the charge fluctuations raises a interesting point.
See Addendum 2 for further discussion.}
It should also be pointed out that the calculation, as presented in
\cite{NEW}, was somewhat model specific. Nevertheless, it is not difficult
to extend this formalism   
 to demonstrate that $b_C=+1$ is ``universal'' in our previously
used sense.
[To put it another way, this result persists
for any (static) black hole
with  a positive  specific heat, provided that  the
system is far from extremality and not too close to a phase-transition point.
These  conditions ensure both
 thermal stability and the leading-order validity of
the critical relation $E\sim A^{\eta}\;$.]

Now, what about the fluctuations due to spin? 
Because  of the  complicated dynamics of a rotating spacetime  
and the vector nature of the angular momentum operator,
such an analysis involves all sorts of technical difficulties.
Nonetheless, we can still reasonably expect that
each  fluctuating degree of freedom will contribute
$+1/2$ to the logarithmic prefactor (as follows from both  the prior
observations and the grand-canonical analysis of \cite{msett}).
This argument leads immediately to a  distinct    
 possibility: Each of the three  spin components 
 (say, $x$, $y$ and  $z$)  fluctuate independently,
in which case the total contribution to the canonical correction
prefactor
is obviously $+3/2\;$. However, this is likely wrong, inasmuch as
we are considering the implications of a quantized area spectrum.
The relevant count should, in all likelihood, be that
of the number of independent quantum numbers required to describe the
spectrum. In the case of a fully general (four-dimensional)
Kerr--Newman black hole, there are compelling reasons to believe
that the number of such  quantum numbers is precisely three;
essentially, one for each of the  mass, charge and spin sectors 
\cite{mak,prior}.~\footnote{Alternatively, one might expect an
equilibrated black hole system to have
its axial asymmetries all smoothed out; thus, favoring
one particular axis of rotation (say, $z$) .}
It then follows that the spin fluctuations, like the charge, will contribute
an additional   amount of $+1/2$ to canonical correction.
Consequently, we arrive at $b_{C}=+3/2$ as our final (canonical) result.
Note that the famous black hole
 ``no-hair'' principle \cite{wheel2} plays an  essential role in restricting
our  considerations to just the three macroscopic contributors: mass,
 charge and spin.  Even if there was some sort of ``exotic (quantum)
 hair'' to 
 account for \cite{col},
it could certainly be dismissed as
inconsequential in  a semiclassical context.

Finally, summing the microcanonical and canonical
contributions, we obtain an  intriguing cancellation:~\footnote{The
possibility of this type of cancellation  has already been
suggested in \cite{zz1}; although on somewhat different grounds.
There, it was pointed out that  isolated-horizon
boundary conditions \cite{zz2} could  reduce the
(loop quantum gravity-based) microcanonical correction
from $-3/2$ to $-1/2$ \cite{zz3} (also see Addendum 1); thus canceling
off the ``universal'' canonical correction of 
$+1/2$.  However, contrary to the current presentation,
this argument completely 
neglects the spin and charge fluctuations.}   
\be
b=b_{mc}+b_{C}=-{3\over 2}+{3\over 2}=0 \;.
\label{8}
\ee

\section{Discussion}

 Our  outcome of $b=0$, although somewhat conjectural, 
does immediately satisfy 
a non-trivial consistency check; namely, 
 it  complies with Hod's statistical  constraint
that the prefactor be  a non-negative integer \cite{hod}.
Moreover, if one interprets the
well-known holographic principle \cite{holo}
in its most literal form (so that $A/4$ is a strict upper bound on 
the black hole entropy),
then  a non-positive value of $b$ would necessarily be favored \cite{majj}.
This realization suggests that {\it zero} is, in fact, 
the {\it unique} choice of prefactor that
is consistent with  both the holographic principle
 and  statistical mechanics.

 However, even if  our arguments 
manage to hold up under closer scrutiny,~\footnote{It should
be noted that there is at least one
rigorous calculation
of the quantum-corrected entropy for which
no logarithmic term appears \cite{huang}.}
 there are (at least) 
a few open questions that should
be addressed.  First
of all, do the lower-order  corrections ({\it i.e.}, ${\cal O}[A^{-1}]$) 
vanish as also predicted by Hod and, perhaps more importantly, 
does the constant term comply with that of equation (\ref{5})? Conversely,
for a macroscopically large  black hole, it is not at all clear 
if  it is sensible
to talk about such small deviations (from a pure integer) in the level 
degeneracy \cite{khrip2}. Secondly, can we expect similar outcomes when the
spacetime dimensionality differs from four? In this regard, it would 
seem reasonable to suggest that, for any viable number of spacetime
 dimensions,
$b$ will  either be a half integer or full integer. However, only
the latter outcome can be deemed as ``acceptable''
(and it would also have to be non-negative). 
At this point in the game, it is difficult to say anything more
conclusive. Undoubtedly,
it will be an instructive exercise
to investigate some specific cases once the analytical ``machinery'' exists. 

To end on  a purely speculative note, it may be of interest
if the value of $b$ did turn out to be some preferred
 non-negative integer, irrespective of the model or number of
dimensions. 
 Since the canonical correction (a {\it quantum} effect) 
is a direct
measure of the variety of ({\it classical}) black hole hair, such an 
eventuality could
be viewed as a signal from the underlying  theory
of quantum gravity. (This type of idea   --- manifestations of quantum gravity 
in the macroscopic world ---  is similar in spirit to
both the holographic principle \cite{holo} and the quasinormal
mode conjecture of \cite{hod2}.)
 Our conjectured  value of
zero would promise even more intrigue; as such a precise cancellation
seems suggestive of some fundamental  symmetry being at play.
Or, perhaps, Nature just abhors a logarithm.


\section*{Addendum 1}

Since the original version of this manuscript, there has been some
dramatic developments in loop quantum gravity. More to the point,
new calculations  of the so-called
Immirzi parameter \cite{Imm} 
have just emerged \cite{Pol-1,Pol-2}. 
As an immediate consequence, it would appear that the  
microcanonical contribution to the 
logarithmic
 prefactor is  $-1/2$ (negating the older  prediction, in loop
quantum gravity,
of $-3/2$).
If this is indeed  correct, then it follows that our net value
for $b$ should be adjusted  from $0$ to $+1$; in which case,
Nature does {\it not} abhor a logarithm after all. Fortunately, this
adjustment
still complies with  Hod's constraint ({\it i.e.}, $b$ should be a
 non-negative integer), so our main point
remains  intact.  Also, it should be kept in mind that the
precise status of these new calculations  
is still a matter of open debate \cite{VN2,VN3}.

\section*{Addendum 2}

Also since the original version of this manuscript,
it has  been suggested \cite{VN} 
that the grand-canonical analysis of \cite{NEW} may be invalidated as follows:
For a classically neutral or almost neutral black hole 
(our regime of interest),
the {\it relative} charge fluctuation ``blows up'' since  
$\Delta Q/Q\rightarrow
\infty$ as $Q\rightarrow 0$. [Actually, the problem occurs because
$<m>\rightarrow 0$, since our partition function (\ref{7}) is
directly expressed in terms of the quantum numbers $n$ and $m$.]
 It is then argued that this divergence
would invalidate the small fluctuation (Gaussian) approximation, which
was used to evaluate the partition function.
We can, however, circumvent this criticism in the following way.
As an alternative to the grand-canonical partition function
 (\ref{7}) used previously, it is just as viable to write
\be
{\cal Z}_G(\beta,L)=\sum_{m_-}\sum_{m_+}\sum_{n}g_{n,m_+,m_-} 
\exp[\beta E(n,m_+,m_-;L)-\mu Q(m_+,m_-)]\;,
\label{X}
\ee
where  $Q=e(m_+-m_-)$ with $m_{\pm}=0,1,2$,...$\;$. 
That is, $m_+$  essentially represents the
number of positively charge particles inside the black hole
and $m_-$, the number of negatively
charged particles. Now, if one proceeds to evaluate
this new version of the partition function (see \cite{NEW}
for the details of the procedure), the result turns out to be identical
to that obtained from equation (\ref{7})  up to a constant 
factor.
(Such a factor is, of course, irrelevant to calculations of the 
canonical entropy, {\it etcetera}.) The important point is that,
for the semiclassical limit of a large black hole, $<m_+>$ and $<m_->$
are both finite (in fact, macroscopically large) irrespective
of whether or not the black hole is, itself,  classically neutral or
charged. Hence, the Gaussian approximation will no longer
break down in the suggested manner.

\section*{Acknowledgments} 

The author would like to thank M. Visser 
for guidance, G. Gour for prior collaborations,
and
S. Hod and P. Majumdar  for  useful correspondences.
 Research is supported  by
the Marsden Fund (c/o the  New Zealand Royal Society) 
and by the University Research  Fund (c/o Victoria University).

\end{document}